\documentclass[11pt]{article}
\usepackage{moriond2000,epsfig}
\usepackage{color} 

\def\Journal#1#2#3#4{{#1} {\bf #2}, #3 (#4)}
\def\ci#1{\hspace{1mm}\cite{#1}}

 \def\rouge#1{\textcolor[named]{Red}{#1}}  
 \def\bleu#1{\textcolor[named]{Blue}{#1}}

\def\mco{\multicolumn}

\begin{document}
\vspace*{4cm}
\title{EROS~2 PROPER MOTION SURVEY FOR HALO WHITE DWARFS}

\author{B. GOLDMAN, for the {\sc Eros} Collaboration }

\address{PCC, Coll{\`e}ge de France, 11 place Marcellin Berthelot, \\
75231 Paris Cedex 05, France}

\maketitle\abstracts{
Since 1996 EROS~2 has surveyed $440^{\circ 2}$ at high Galactic latitude 
to search for high proper motion stars in the Solar
neighbourhood. We present here the analysis of $250^{\circ 2}$ for
which we have 
three years of data. No object with halo-like
kinematics has been detected. Using a detailed Monte-Carlo simulation
of the observations, we calculate our 
detection efficiency 
for this kind of object and place constraints on their contribution
to various halo models. If 14~Gyr old, the halo cannot be made of more
than 18\% 
of hydrogen white dwarfs (95\% C.L.).
}

\section{Introduction}

Cool white dwarfs (WDs) have become a very popular subject 
in the recent years, since
the {\sc Macho}\ci{Al97} results suggested that they contribute to
the halo 
introduced to explain the Galactic rotation curve. 
Recent atmosphere models\ci{Sa99} taking into account collision induced
absorption predict the hydrogen WDs to be bluer than earlier,
reducing the previous constraints from colour- and infrared surveys.
Such blue cool WDs have recently been discovered\ci{Ib00}.

On the other hand, the {\sc Eros} collaboration\ci{La00}
lowered its limit on the microlensing optical depth 
towards the Magellanic Clouds.
A large number of population III WDs in the halo
would also contradict some of our current ideas:
the strange IMF of their progenitors,
metal and Helium enrichment of the Galaxy,
extragalactic observations of young haloes
and observations of multi-TeV gamma rays\ci{Gr99}.
%
Hence the question remains to know what the {\sc Macho} lenses are.
Cool WDs will also teach us about the early stages of star formation
in the Galaxy and its age, 
and any WD older than those known today would bring important informations.

Proper motion surveys are a way to distinguish cool WDs from the
more numerous, brighter and more distant disk stars
by means of their much higher proper motion.
Thus, EROS started a large survey, and we report the results here.  
We first describe the data we used and the way the high proper motion
catalogue was created; then we present the current results of the 
EROS~2 survey about the contribution of cool hydrogen white dwarfs to 
the halo.

\section{Description of the Survey}

\subsection{The Data}
We used the {\sc Eros}~2 wide field imager, located at La~Silla Observatory,
Chile. The instrument takes two $1^{\circ 2}$, 
$8k\times 4k$ CCD images simultaneously in two broad band
filters, a visible band, between $V$ and $R$, and a red band close to 
$I$. Observations for the proper motion survey are conducted nearly
all year round, during dark time, close to the meridian to reduce 
atmospheric refraction. Exposure time varies between 5 and 10~minutes, 
with limiting magnitudes of 
\hspace{1cm}
$V\simeq\,21.5$ in the visible band and 
$I\simeq\,20.5$ in the red band.
The pixel size is 0.''6.

Since 1996 we have taken
3,600 images 
of 442 
fields located at high Galactic latitudes, mostly
at 22.5h$<\alpha<3.5$h and $\delta=-39^\circ,-45^\circ$ 
(South Galactic Pole fields),
and at 10h$<\alpha<14.4$h and $\delta=-6^\circ,-12^\circ$ 
(North Galactic Hemisphere fields). 
In this paper we use data taken between 1996 and
1999, for $250^{\circ 2}$ 
for which we have three epochs, separated by one year.
The remaining fields will be analyzed when we get a third epoch.

\subsection{The Reduction}

The flat-fielding and debiasing were performed using the {\sc Eros} 
package {\sc Peida}. The astrometric reduction was done by fitting 
a two dimensional Gaussian on the objects detected by correlation 
with a Gaussian PSF. The parameters of the fit were used
in order to remove non-stellar objects 
(bad pixels, cosmic rays and the largest galaxies).
The images were then geometrically aligned over $11\,$arcmin chunks,
using the bright stars of the field, as galaxies are not numerous
enough nor well measured.
Then the stars were matched with a search radius corresponding to
a maximum proper motion of $6''/$yr, requiring that the fluxes be
compatible within $4\sigma$. 
We produce this way multi-epoch catalogues in two bands.

\begin{figure}[htb]
\begin{center}
\mbox{
\epsfig{figure=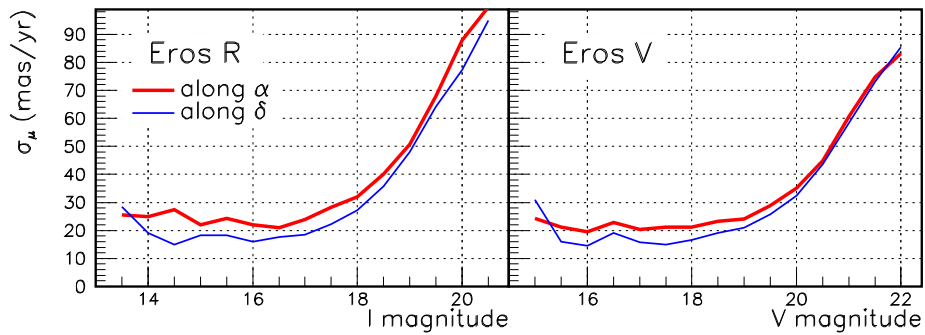,height=44mm} } 
\caption{Proper motion dispersion for Eros R and Eros V images,
 along $\alpha$ (thick lines) and $\delta$ (thin lines).
 \label{fig:erreurs} }
\end{center}
\end{figure}

Errors were described by photon statistics and 
by the dispersion of bright star positions.
The first contribution dominates for the faint stars, 
in which we expect most of our halo candidates.
The second is due to the optical
deformation of the (wide-field, focal reduced) telescope and the 
proper motion of the stars used to determine our reference frame. 
Total errors on a single frame range from 30~mas for bright objects, 
up to 150~mas at the detection limit.
The external errors on the proper motion measurements are presented 
in Fig.\ref{fig:erreurs}.

\subsection{The Selection Criteria}
To eliminate the usual contamination by asteroids, noise detection, 
remaining cosmic rays and galaxies, we require three
detections, among a set of 3 to 8 images, over three years. 
We then impose that the confidence level
of the proper motion fit along $\alpha$ and $\delta$ be higher
than 0.5\%. At this point, depending on the population we
want to select, we apply different sets of cuts:

\noindent $\bullet$ {\bf slow populations}: 
objects from the 
disk are intrinsically slow, with a velocity dispersion of 
$\sigma\simeq\,20-50\,km/s$ depending on the age of the stars. At
a distance of 100~pc, this translates to a proper motion of 
$\mu = \frac{V_{\bot}\;(km/s)}{4.74\,D\;(pc)} \approx\,80\,$mas/yr 
which is only a $\sim\,2-\sigma$ detection
even for 
bright stars. This means that no proper motion will be measurable for
faint stars over our small time baseline, 
except for the fastest or closest ones
(see ~\ref{subsec:resdisk} for an example), and that
a cross-selection between the two band catalogues will be needed, to
remove noise contamination (bicolour analysis). 
Thus we require that the proper motion be
higher than $60\,$mas/yr and $3.5\sigma$, 
and that the visible and red directions 
of proper motion be within $40^\circ$. 
This selects stars brighter than 
18, as fainter stars have too large proper motion errors to be selected. 

\noindent $\bullet$ {\bf halo}: 
here we expect proper motions of 1''/yr or
more, which makes any detection very significant, even with  one 
single band (monochrome analysis),
and intrinsically faint stars of $M_{V(I)}>16(15)$. 
Thus to remove the slower, brighter known stars
we require that 
the reduced proper motion (RPM)
$H_{V}=M_{V}+5log(V_{\bot})-3.378=V+5log(\mu)+5$,
where $V_{\bot}$ is the transerve velocity in km/s,
be higher than 21 in the visible band, and $H_{I}>20$ in the red band.
Any star with $V>16$ or $I>15$ and $V_{\bot}>50$km/s will satisfy this 
cut.
Additionally, as a $V=21$ disk star may have a 3--$\sigma_{\mu}$ spurious proper motion of 
0.4''/yr and a misleading $H_{V}=24$, we also require 
the proper motion 
be higher than
0.7''/yr, or 200~km/s at 60~pc. 
This cut removes between 30\% ($M_{V}=16.5$, co-rotation of 50~km/s)
and 10\% ($M_{V}=18$, no rotation) of detectable halo stars.

\section{Results}

\subsection{Candidates of known Populations}\label{subsec:resdisk}

Following the steps described above we select 
1,046 objects 
in the visible band 
and 1,079 
in the red one.
Careful examination of these objects
reveal that the sample is not free from
contamination by spurious detections.
Most candidates have a 
reduced proper motion 
$H_V$ between 17 and 22. 
Objects bluer than $V-I\simeq\,1.2$ can be
interpreted mainly as thin and thick disks white dwarfs, 
and redder objects as disks red dwarfs. 
Some candidates with higher reduced proper motion, which may be
spheroid objects or nearby, very cool dwarfs have been 
followed up spectroscopically. 
{\sc Denis}\ci{De99} photometry has been checked for the reddest
candidates. For example, {\sc Lhs102b}\ci{BG99}
was first detected as a high proper 
motion star, then associated with {\sc Lhs102} 
through their common proper motion\ci{Yale}, 
and finally confirmed as a L--dwarf by {\sc Denis} and
spectroscopy. 

\begin{figure}[htb]
\begin{center}
\mbox{
\epsfig{figure=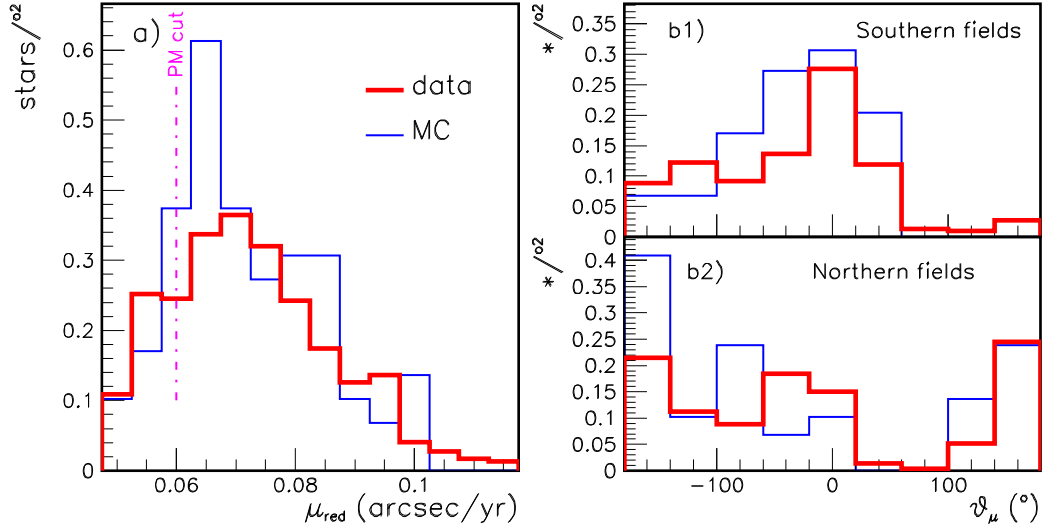,height=70mm} } 
\caption{
 Distribution of 
 detected (thick line) and 
 expected (thin line) bright ($14<I<16.6$) bicolour candidates,
 for the red band.
 Expectations from a Besan{\c c}on simulation of the Galaxy,
 corrected for our detection efficiency.
 a) Proper motion distribution.
 b) Proper motion direction distribution in our South Galactic Pole 
 fields (top) and our Northern Galactic Hemisphere fields (bottom).
 \label{fig:mu+dir} }
\end{center}
\end{figure}

On a statistical point of view, we might check whether we reproduce well 
our efficiencies and our errors 
by comparing the thin
and thick disks' and spheroid prediction with our data set. 
For this we use the Besan{\c c}on
model of the Galaxy\ci{Ro86} to create artificial catalogues of
stars in our fields, then simulate the observations for these stars,
using the actual distribution of each field (atmospheric conditions,
number of exposures,...).
We compare star counts, which indicate compatibility
between the data and the Monte-Carlo for bright stars, at the 15\% 
level. 
For faint objects the stars are dominated by galaxies and detailed
comparison is impossible until we get a better classification of our
objects.
We can then apply to the resulting simulated data
the {\em slow populations} set of cuts to check our sensitivity to
{\em small} proper motions. 
We observe in Fig.\ref{fig:mu+dir}a that 
the results agree with the model within 20\%.
As we lack detailed explanation for the difference,
we conservatively lower our sensivity by 20\%.
We also checked that our proper motion direction distribution,
which mainly depends on the Sun own peculiar motion,
agrees well with the Besan{\c c}on distribution
(see Fig.\ref{fig:mu+dir}b).

\subsection{Constraints on the Halo}

\begin{figure}[htb]
\begin{center}
\mbox{
\epsfig{figure=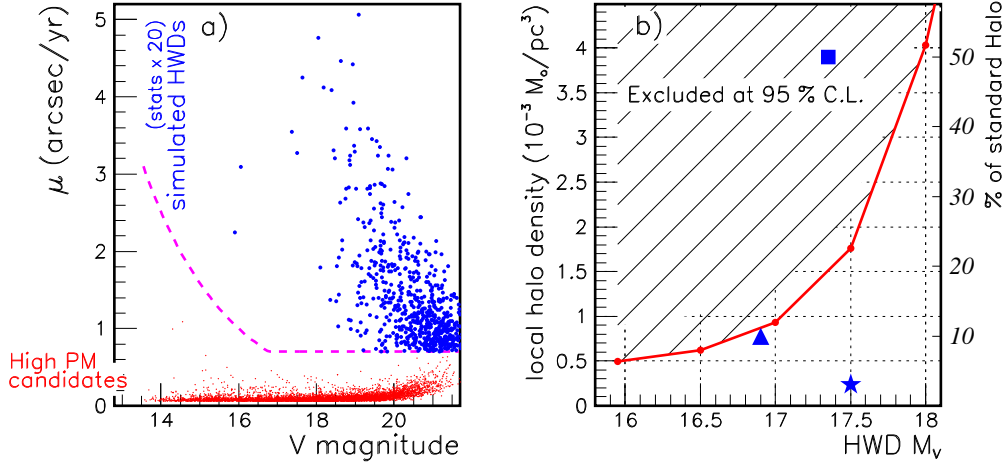,height=62mm} } 
\caption{
 a) Proper motion vs. $V$ for high PM monochrome candidates (dots) 
 and simulated 
 $M_V=17$,
 $0.6M_\odot$ HWDs passing the cuts 
 (statistics $\times 20$, circles), for the visible band.
 The halo PM cut is indicated by the dashed line. 
 b) Exclusion diagram as a function of the halo WD $M_V$ magnitude,
 for $0.6M_\odot$ HWDs.
 The WD halo fraction central values suggested by
 Ibata {\it et~al}$\,^8$ (square),   
 Ibata {\it et~al}$\,^9$ (triangle), 
 Flynn {\it et~al}$\,^5$ (star)      
 are indicated.
 \label{fig:halo+excl} }
\end{center}
\end{figure}

As with the known populations simulation, we simulated various kinds of 
haloes made of old hydrogen white dwarfs, with different HWD ages, masses and
kinematics.
The luminosity function of HWDs and the colour--magnitude
function are those 
of Chabrier\ci{Cha99} and Saumon \& Jacobson~\cite{Sa99}. 
We also indicate our
sensitivity to haloes with Dirac-like luminosity functions
(see Table~\ref{tab:expres}).
The detailed 
distribution of magnitudes is not crucial as only the stars in the 
brighter part of the luminosity function 
($M_V=17.2\pm0.2$ for a 14~Gyr halo, $M_{V}\approx 17.8\pm0.3$ for 15~Gyr) 
contribute. Colour is important as a $V-I < 0.5$ WD will be easily 
missed in the red band, or a $V-I > 1.5$ in the visible, but this 
effect is compensated in our survey by the independent use of the two filters.
Finally the cooling curve evolves with the WD mass; 
$1.2M_\odot$~WDs cool faster than $0.6M_\odot$~ones\ci{Cha00}.
One should also remember that we measure a local {\em number} density 
rather than a mass density.
The kinematics used correspond to the usual sets of parameters 
observed for the spheroid and expected for the halo, with a small 
rotation $-50<\omega<50$~km/s and velocity dispersion of 
150--250~km/s. This is not crucial as most of the nearby HWDs in any model 
will have a proper motion above the 0.7''/yr cut.
Conservatively we apply the correction for our sensitivity mentioned 
above, although it concerns the bicolour search for small proper 
motions, while we search here for large proper motion in each band 
independently. 
An example of a simulation is shown Fig.\ref{fig:halo+excl}a.

We find no halo type candidates in any band. 
Given our expectations of 
Table~\ref{tab:expres}, we exclude a local number density 
higher than 
$2.3\,10^{-3}\,$WD$/pc^3$
of 14~Gyr HWDs  
at the 95\% C.L. 
The exclusion diagram, as a function of the halo WD $M_V$ magnitude,
is shown in Fig.\ref{fig:halo+excl}b.

\begin{table}[t]
 \caption{Expectation and constraints for HWDs
  of a certain $M_V$ magnitude, and for different halo ages. 
  Density limit at 95\% C.L.
  For the expectation and mass density figures,
  we assumed a local density of
  $7.8\,10^{-3}M_{\odot}/pc^3$, for $0.6M_\odot$~HWDs.
  Figures in the left column are for the red band,
  in the right band for the visible band.
  \label{tab:expres}}
 \vspace{0.4cm}
 \begin{center}
 \begin{tabular}{|r|c|c|c|c|c|c||c|c|c|c|l|}
  \hline
   & \mco{6}{|c||}{$M_V$ magnitude of HWDs} &
   \mco{4}{|c|}{Halo age (Gyr)} & \\
  \cline{2-11}
   & \mco{2}{|c|}{16.5} & \mco{2}{|c|}{17} & \mco{2}{|c||}{17.5} & 
     \mco{2}{|c|}{14} & \mco{2}{|c|}{15} & \\
  \hline
  explored vol. &
  \rouge{2.5} & 
  \bleu{2.9} & 
  \rouge{1.5} & 
  \bleu{1.9} & 
  \rouge{0.5} & 
  \bleu{1.0} & 
        
  \rouge{0.9} & 
  \bleu{1.3} & 
  \rouge{-} & 
  \bleu{0.3} & 
  1000~pc$^3$\\
  \hline
  expectation &
  \rouge{33.1} & 
  \bleu{37.6} & 
  \rouge{19.2} & 
  \bleu{25.1} & 
  \rouge{6.7} & 
  \bleu{13.3} & 
        
  \rouge{11.4} & 
  \bleu{17.2} & 
  \rouge{0.7} & 
  \bleu{3.6} & 
  WDs\\
  \hline
  star density & 
  \rouge{1.2} &  
  \bleu{1.0} &  
  \rouge{2.0} &  
  \bleu{1.6} &  
  \rouge{5.8} & 
  \bleu{2.9} & 
  
  \rouge{3.4} &  
  \bleu{2.3} &  
  \rouge{-} & 
  \bleu{11} & 
  $10^{-3}\,$WD/$pc^3$ \\
  \hline
  mass density & 
  \rouge{0.7} &  
  \bleu{0.6} &  
  \rouge{1.2} &  
  \bleu{0.9} &  
  \rouge{3.5} & 
  \bleu{1.8} & 
  
  \rouge{2.0} & 
  \bleu{1.4} &  
  \rouge{-} & 
  \bleu{6.6} &  
  $10^{-3}\,M_\odot/pc^3$ \\
  \hline
 \end{tabular}
 \end{center}
\end{table}

\section{Conclusion}
Using the first three years of data for 282 EROS~2 fields, we can
place an upper limit of 18\% 
(95\% C.L.)
to the contribution of white
dwarfs of age 14~Gyr or of magnitude $M_{V}=17.2$.
We are compatible with the EROS microlensing results towards the Magellanic
Clouds\ci{La00} and the analysis of proper motion observations 
by Flynn~\&~al\ci{Fl99} and Ibata~\&~al\ci{Ib00}. 
We exclude a 50\% contribution by $M_V\approx 17$ HWDs suggested by 
Ibata~\&~al\ci{Ib99} or Mendez~\&~al\ci{Me00},
whose blue point-like objects remain to be identified.
We do not place constraints on fainter objects, either older
or with a pure-helium atmosphere. 
In the coming months additional data will become available, so that
more fields will be analyzed in addition to those presented here.

\vspace{.4cm}
\noindent{\em Note:}
During the talk in Les~Arcs we presented the results of the bicolour 
halo analysis.
Since then we progressed 
on the more sensitive, monochrome halo analysis, which is presented here.

\end{document}